\documentclass[conference]{IEEEtran}
\IEEEoverridecommandlockouts
\usepackage{graphicx}
\usepackage{booktabs}
\usepackage{array}
\usepackage{amsmath,amssymb}
\usepackage{url}
\usepackage[hidelinks]{hyperref}
\usepackage{balance}
\usepackage{listings}
\providecommand{\passthrough}[1]{#1}
\providecommand{\tightlist}{\setlength{\itemsep}{0pt}\setlength{\parskip}{0pt}}
\begin{document}
\title{Don't Let the Model Write the YAML: Deterministic, Minimal-Diff GitOps Remediation from LLM-Proposed Field Changes}
\author{\IEEEauthorblockN{Pruthvi Davineni}
\IEEEauthorblockA{Independent Researcher\\ pruthvi.davineni@gmail.com}}
\maketitle
\begin{abstract}
LLM agents increasingly diagnose production incidents and propose remediations. In a GitOps workflow, applying such a fix means editing a version-controlled configuration file. The obvious implementation, having the model author the edited file or a diff of it, is the one practitioners reach for first. We evaluate that choice on real Kubernetes manifests and find that \textbf{which failure you get depends on how the model is asked and how capable it is, but that no text-generation strategy is suitable for unattended automation.} Asking a model for a \textbf{unified diff} is unsafe: under strict patching almost none apply, but that is an artifact: a tolerant patch tool (GNU \texttt{patch\ -\/-fuzz}) applies 96\% of the same diffs, yet \textbf{silently misapplies \textasciitilde1 in 7} (14--20\%), landing the change at the wrong place with no error signal. Asking a model to \textbf{rewrite the whole file} is capability-dependent: a small model reflows and corrupts the file, while a frontier model is \emph{usually correct}, but is \textbf{non-deterministic} (the same request silently drops a field or edits a neighbor on a fraction of tasks) and must regenerate the entire file, so its cost is \emph{O(file size)} per edit no matter how small the change.

We present an alternative that \textbf{separates the semantic decision (which resource, which field, which value) from the syntactic act of editing the file.} The agent emits only a structured \emph{field-change intent}; a deterministic pipeline indexes manifests by Kubernetes \texttt{(kind,\ name)}, locates the target scalar's exact character span using the YAML parser's node position marks, and replaces \emph{only} that span in the raw text. Because the file is never re-serialized, the diff is minimal by construction (a single scalar), formatting and comments are preserved, and the edit is \textbf{correct and deterministic independent of the model}: the same result whether driven by a small model or a frontier one, at \emph{O(1)} generation cost (a fixed field-change intent, independent of file size) rather than the \emph{O(file size)} cost of regenerating the file. The contribution is not the span-edit primitive (structured editing is well known) but the pairing of an LLM-proposed intent with a deterministic, fail-closed application contract for GitOps. We implement it in KubeAstra (Apache-2.0) and release the benchmark (corpus, oracle, harness). Our claim is scoped to \emph{faithful application} of a known change; whether the change is the \emph{right} one is left to human PR review.

\begin{center}\rule{0.5\linewidth}{0.5pt}\end{center}

\end{abstract}

\subsection*{1. Introduction}\label{introduction}
\addcontentsline{toc}{subsection}{1. Introduction}

LLM agents now propose infrastructure fixes, and GitOps (Argo CD, Flux) {[}10, 11, 12{]} makes the Git repository the source of truth: to ``apply the fix'' is to \emph{edit a manifest and open a pull request}. The na\"ive path, where the model writes the new file or a diff, is the one everyone reaches for, and, as we show, the one that fails or hides a latent hazard.

This paper makes a narrow, testable claim about the \textbf{application} step, deliberately separated from the (harder, human-reviewed) question of whether the proposed fix is \emph{correct}. Given a known, correct semantic change (``\texttt{api-gateway} needs \texttt{spec.replicas:\ 5}''), how should the system turn it into bytes on disk? We compare three text-generation strategies against a deterministic one, across a small and a frontier model, on real manifests.

\textbf{Findings (previewed):}

\begin{enumerate}
\def\labelenumi{\arabic{enumi}.}
\tightlist
\item
  \textbf{Diff generation is unsafe for unattended use, and the reason is subtle (\textsection{}6.7).} Under \emph{strict} application almost no model diffs apply (2.7\% for the frontier model), but that is a lower bound: most rejections are wrong \texttt{@@} line numbers, not wrong content, and a tolerant tool (GNU \texttt{patch}) applies 96\%. The real problem is that \textbf{14--20\% of the \emph{applied} diffs are silently misapplied} (wrong result, no error), which in an automated commit-and-ship pipeline is worse than an honest rejection. A validate-and-retry loop does not fix it (the retry re-authors a diff with the same failure mode).
\item
  \textbf{Full-file rewrite is capability-dependent and unsafe for automation.} A small model reflows the file wholesale (2.4\% correct; 97.6\% of outputs alter an unrelated line; mean diff 921 lines). A frontier model rewrites faithfully most of the time (97.6\% correct, near-minimal diff), but is \textbf{non-deterministic}: 7.2\% of tasks (6/83) yield a correct edit on some seeds and a corrupted one (dropped field, collateral change) on others. For a change that will be committed and shipped, \emph{non-deterministic silent corruption} is disqualifying.
\item
  \textbf{Deterministic span-editing removes the model from byte-level editing entirely.} It is 100\% correct, 0\% collateral, perfectly minimal (one scalar), and \textbf{identical across both models and all 415 runs}, because the model never authors file content.
\end{enumerate}

\textbf{Thesis.} The model is good at the \emph{semantic} question and unreliable at the \emph{syntactic} one. Constrain the model to a structured intent; apply it deterministically. Correctness then follows from the algorithm rather than the model's capability, and is free of the cost, latency, and variance of regenerating file text.

\subsubsection*{Contributions}\label{contributions}
\addcontentsline{toc}{subsubsection}{Contributions}

\begin{itemize}
\tightlist
\item
  \textbf{C1.} A design principle for safe LLM-driven configuration change: \emph{separate the semantic intent (resource, field, value) from the deterministic application to the file}, with the resulting minimal-diff, no-fabrication, and determinism guarantees stated and proven by construction.
\item
  \textbf{C2.} A concrete mechanism, \textbf{YAML node-mark span-editing}: replace exactly the target scalar in the raw text using the parser's character-position marks, preserving comments, quoting, and key order without re-serialization; with named-list resolution (containers/env by name, not index) and explicit refusal on ambiguous or absent targets.
\item
  \textbf{C3.} An open-source implementation in KubeAstra (Apache-2.0): repository indexing without a \texttt{git} binary or disk writes, a two-phase preview$\rightarrow$open protocol, a CI-enforced no-auto-merge invariant, and a Kustomize patch-append fallback.
\item
  \textbf{C4.} A reproducible benchmark and evaluation on real manifests across two models and five seeds, measuring correctness, collateral change, fabrication, format preservation, diff minimality, and \textbf{determinism}, the last being where the deterministic approach separates decisively from a frontier model that otherwise looks competitive.
\end{itemize}

\subsection*{2. Background \& Motivation}\label{background-motivation}
\addcontentsline{toc}{subsection}{2. Background \& Motivation}

\textbf{GitOps.} Desired state lives in Git; a controller reconciles the cluster to it; changes flow through reviewed PRs, giving audit, rollback, and a human gate {[}10{]}.

\textbf{LLM agents for operations.} ReAct-style agents {[}1{]} have moved from \emph{suggesting a command} to \emph{changing the configuration}. The remediation step is where an agent stops advising and starts editing production source.

\textbf{Why YAML is hostile to ``just re-dump it''.} Kubernetes YAML is whitespace-significant, comment-bearing, supports anchors/aliases and multi-document files, and encodes operator intent in comments and key ordering. Parsing to an object model and re-serializing is lossy: it reflows, re-quotes, reorders, and strips comments. Any approach that round-trips the file through a serializer forfeits fidelity before the model is even involved.

\textbf{The gap.} Existing agent tooling either (a) applies changes imperatively to the live cluster, bypassing GitOps, or (b) lets the model author the file or diff, which we show is unreliable. Neither yields a reviewable, minimal, guaranteed-faithful GitOps change.

\subsection*{3. Problem Statement \& Design Goals}\label{problem-statement-design-goals}
\addcontentsline{toc}{subsection}{3. Problem Statement \& Design Goals}

\textbf{Setting / threat model.} A semi-autonomous agent proposes a change; a human reviews the PR; the agent is \emph{not} trusted to write file bytes or to merge. Even a well-intentioned model is an untrusted diff author.

\textbf{Design goals (testable).}
- \textbf{G1 Minimal diff:} the change touches only the intended scalar; diff size is bounded by the edit, not the file.
- \textbf{G2 No fabrication / no collateral:} every byte outside the target is unchanged.
- \textbf{G3 Fidelity:} comments, quoting style, key order preserved.
- \textbf{G4 Determinism:} identical intent + repo state $\Rightarrow$ identical diff.
- \textbf{G5 Fail-closed:} ambiguous or absent target $\Rightarrow$ refuse, never guess.
- \textbf{G6 Reviewability:} output is a PR; a human merges; the agent cannot.

\textbf{Scope (v1).} Scalar field changes (\texttt{replicas}, container image tag, resource requests/limits, env values) on \texttt{Deployment}/\texttt{StatefulSet}-style resources, in plain-YAML and Kustomize repositories. Non-goals (Helm/Argo value indirection, structural edits) are stated in \textsection{}7.

\subsection*{4. Approach}\label{approach}
\addcontentsline{toc}{subsection}{4. Approach}

\subsubsection*{4.1 The field-change intent (the interface --- C1)}\label{the-field-change-intent-the-interface-c1}
\addcontentsline{toc}{subsubsection}{4.1 The field-change intent (the interface --- C1)}

The agent emits a structured intent, not file text:
\texttt{FieldChange\{\ kind,\ name,\ namespace?,\ field\_path,\ new\_value,\ reason\ \}}.
\texttt{field\_path} uses map keys with \textbf{named} list segments (a container by its \texttt{name}, an env var by its name), never positional indices. This is the crux: the model's entire contribution is \emph{which scalar and what value}.

\subsubsection*{4.2 Repository indexing without git or disk}\label{repository-indexing-without-git-or-disk}
\addcontentsline{toc}{subsubsection}{4.2 Repository indexing without git or disk}

Fetch the repo as a gzip tarball via the hosting API; extract in memory; parse every \texttt{*.y*ml} with a safe multi-document loader; \textbf{skip files that don't parse} (Go-templated Helm charts) rather than failing. Index each document by \texttt{(kind,\ name)}; namespace disambiguates rather than keys (Kustomize sets it centrally). No \texttt{git} binary, no token on disk, no manifest cache.

\subsubsection*{4.3 Node-mark span location (the mechanism --- C2)}\label{node-mark-span-location-the-mechanism-c2}
\addcontentsline{toc}{subsubsection}{4.3 Node-mark span location (the mechanism --- C2)}

Span-editing via a parser's position marks is a \emph{known} structured-editing technique (IDE refactorings, tree-sitter tooling {[}7{]}, and semantic patchers {[}3{]} all edit located spans); our contribution is not the primitive but its use under an LLM-intent + fail-closed refusal contract for GitOps remediation. Compose the target document into a node tree whose scalar nodes carry \texttt{start}/\texttt{end} character-position marks. Walk the tree along \texttt{field\_path}, resolving named-list segments by matching the item's \texttt{name} child, and return the terminal scalar's \texttt{{[}start,\ end)} character span.

\begin{figure*}[t]
\centering
\includegraphics[width=0.82\textwidth]{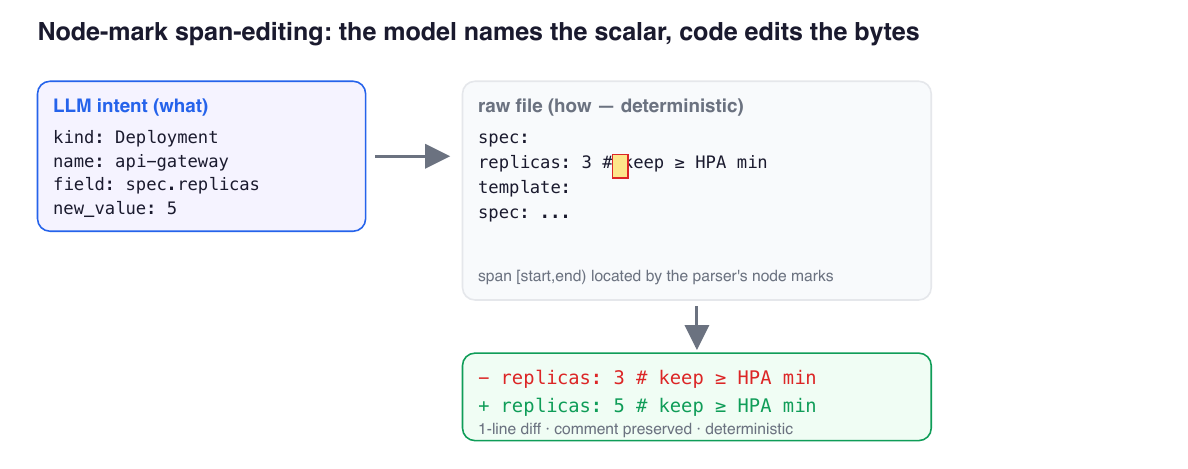}\\[3pt]
{\footnotesize \textbf{Figure 1:} node-mark span-editing --- the model names the scalar; deterministic code locates its byte span and replaces only it, yielding a 1-line diff with the comment preserved.}
\end{figure*}

\textbf{Ambiguity (G5):} 0 matches $\Rightarrow$ refuse (hinting if Helm/Argo markers are present); \textgreater1 $\Rightarrow$ prefer the environment overlay, else refuse and list candidates; field absent $\Rightarrow$ refuse.

\subsubsection*{4.4 Span replacement (why the guarantees hold --- C1)}\label{span-replacement-why-the-guarantees-hold-c1}
\addcontentsline{toc}{subsubsection}{4.4 Span replacement (why the guarantees hold --- C1)}

\texttt{edited\ =\ raw{[}:start{]}\ +\ requote(new\_value)\ +\ raw{[}end:{]}}, with no re-serialization. \texttt{requote} reproduces the original scalar's quoting style so a quoted numeric stays quoted (avoiding a type coercion and a larger diff).
- \textbf{G1/G2:} every byte outside \texttt{{[}start,end)} is unchanged by construction $\Rightarrow$ minimal diff, zero collateral, zero fabrication.
- \textbf{G3:} comments/quoting/order live outside the span $\Rightarrow$ preserved.
- \textbf{G4:} pure function of (raw text, span, value) $\Rightarrow$ deterministic.

\textbf{Lemma (minimality).} Let the located span be \texttt{{[}s,e)} and \texttt{r\ =\ requote(v)}; the output is \texttt{raw{[}:s{]}\ +\ r\ +\ raw{[}e:{]}}. Every byte outside \texttt{{[}s,e)} is copied verbatim, so the line-diff against \texttt{raw} touches only the line(s) spanning \texttt{{[}s,e)}. For the in-scope field types a scalar span lies within one line, so the diff is exactly that one line, and its edit distance equals the edit distance between the old and new scalar representations, the minimum possible for the change. Collateral and fabrication are identically zero. \emph{(The guarantee is over the byte-level edit; it does not assert the change is semantically correct; that is the intent's responsibility, \textsection{}7.)}

\subsubsection*{4.5 Kustomize patch-append fallback}\label{kustomize-patch-append-fallback}
\addcontentsline{toc}{subsubsection}{4.5 Kustomize patch-append fallback}

When the resource is generated (no checked-in source line) but the repo is Kustomize with a known overlay, append a strategic-merge patch and its \texttt{kustomization.yaml} entry: one commit, still reviewable, the only case that adds a file.

\subsubsection*{4.6 From edit to pull request (C3)}\label{from-edit-to-pull-request-c3}
\addcontentsline{toc}{subsubsection}{4.6 From edit to pull request (C3)}

\textbf{Two-phase protocol:} \texttt{preview} returns the real diff and a single-use token (nothing pushed); \texttt{open} pushes branch+commit+PR via the hosting Git Data API (blob$\rightarrow$tree$\rightarrow$commit$\rightarrow$ref$\rightarrow$pull). \textbf{No-merge invariant:} no code path calls a merge endpoint, enforced by a CI test. The human always merges.

\subsection*{5. Implementation}\label{implementation}
\addcontentsline{toc}{subsection}{5. Implementation}

KubeAstra (Apache-2.0). The GitOps component is \textasciitilde520 lines of Python across eight modules; the safety-critical core, the locator (\textsection{}4.3) and the span editor (\textsection{}4.4), is \textasciitilde150 lines. Zero new native dependencies: an HTTP client and the stdlib-level YAML parser only; heavier libraries were avoided so the signed, notarized desktop distribution stays simple. 36 dedicated tests cover the locator, round-trip invariants, quote preservation, and the CI-enforced no-merge guard. The component runs identically in server and desktop modes; the token comes from an environment secret or the OS keychain.

\emph{(The mechanism's small size is itself a result: correctness comes from \textasciitilde150 lines of deterministic code, not from a capable model.)}

\subsection*{6. Evaluation}\label{evaluation}
\addcontentsline{toc}{subsection}{6. Evaluation}

\subsubsection*{6.1 Research questions}\label{research-questions}
\addcontentsline{toc}{subsubsection}{6.1 Research questions}

\begin{itemize}
\tightlist
\item
  \textbf{RQ1 (correctness):} Given a known change, does each strategy produce \emph{exactly} that change?
\item
  \textbf{RQ2 (safety, collateral/fabrication):} Does it alter or invent anything else?
\item
  \textbf{RQ3 (fidelity \& minimality):} Are comments/formatting preserved and the diff minimal?
\item
  \textbf{RQ4 (determinism):} Does the same request produce the same edit across seeds?
\item
  \textbf{RQ5 (capability dependence):} How do the answers change between a small and a frontier model?
\item
  \textbf{RQ6 (fail-closed, G5):} On adversarial targets (absent, ambiguous, templated, non-scalar, alias), does the locator refuse rather than guess?
\end{itemize}

\subsubsection*{6.2 Corpus}\label{corpus}
\addcontentsline{toc}{subsubsection}{6.2 Corpus}

83 field-change tasks generated from two Apache-2.0 sources, SHA-pinned with recorded provenance: the \textbf{Online Boutique} manifests (a single \textasciitilde9.3K-token, multi-resource file, the realistic hard case) and a Kustomize \textbf{helloWorld} base/overlay. Each task is \texttt{(manifest,\ kind,\ name,\ field\_path,\ old\_value,\ new\_value)} with a deterministic new value; field types span replicas, image tag, resource requests/limits, and env values.

\textbf{We are explicit about the corpus's limits.} Because the multi-resource file yields many edit sites, \textbf{\textasciitilde80 of the 83 tasks are edits within that single file}; the corpus thus exercises many \emph{edit sites} but few \emph{distinct files}. Reported rates should be read as edit-site rates within a small file set, \textbf{not population estimates}; we therefore round to two decimals and treat single-file-dependent findings (notably the flaky-6 determinism result) as indicative, to be confirmed on the $\geq$250-task, multi-repo corpus planned for the peer-reviewed version (\textsection{}7). This preprint's claims rest on effects that are large and consistent across all 415 runs (diffs failing under strict patching; span-edit's by-construction determinism), not on the third decimal of any rate.

\subsubsection*{6.3 Systems}\label{systems}
\addcontentsline{toc}{subsubsection}{6.3 Systems}

\begin{itemize}
\tightlist
\item
  \textbf{SUT:} the deterministic span-edit pipeline (\textsection{}4), given the structured intent.
\item
  \textbf{B1 (full-file):} the model rewrites the entire edited file.
\item
  \textbf{B2 (unified-diff):} the model emits a unified diff, applied strictly (any context mismatch = failure, a real, measured weakness, not an artifact).
\item
  \textbf{B3 (diff + retry):} B2 plus validate-and-retry (re-prompt with the apply/parse error, up to 2 retries).
\end{itemize}

All baselines receive the same engineered prompt with a natural-language change description and a worked one-shot example (a \emph{fair}, not straw-man, comparison; prompts in Appendix A). Two models: \textbf{\texttt{gemini-3.7-flash}} (small) and \textbf{\texttt{claude-sonnet-5}} (frontier). \textbf{5 seeds}, 415 runs per system per model. Correctness is judged by an automated \textbf{parse-and-compare oracle}: it parses both manifests, confirms the target field equals \texttt{new\_value}, and flags any \emph{other} changed field as collateral and any added structure as fabrication, so a semantically-correct-but-reflowed output still scores correct on RQ1 while failing fidelity, and baselines are never straw-manned on formatting alone.

\subsubsection*{6.4 Results}\label{results}
\addcontentsline{toc}{subsubsection}{6.4 Results}

\begin{figure*}[t]
\centering
\includegraphics[width=0.82\textwidth]{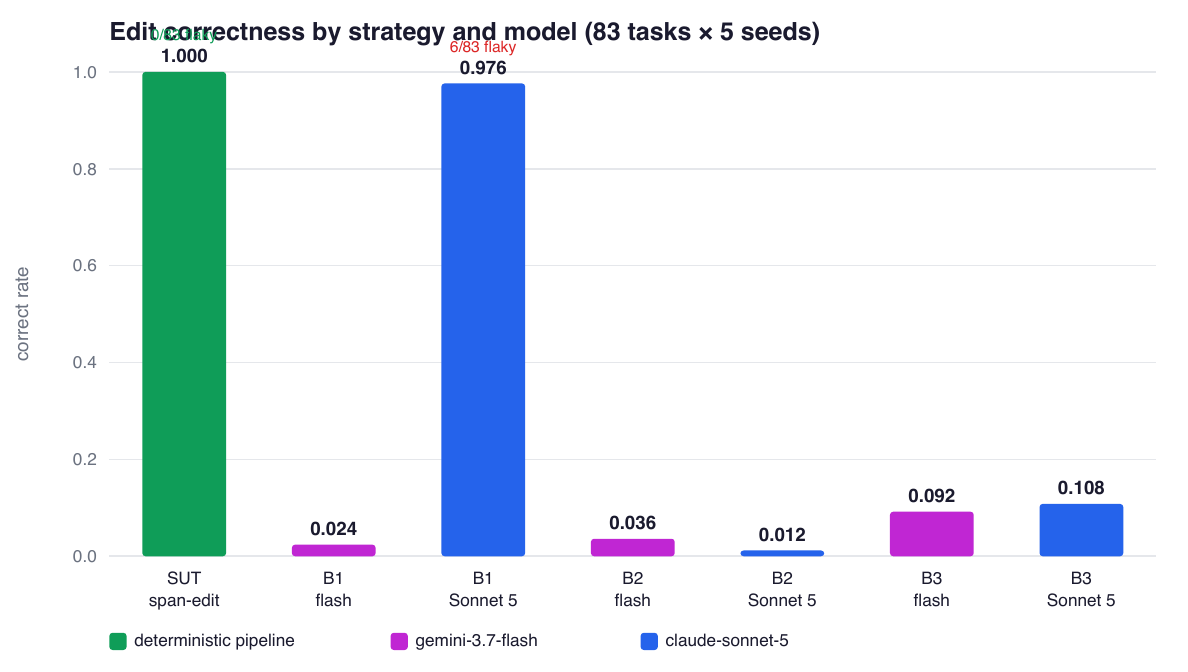}\\[3pt]
{\footnotesize \textbf{Figure 2:} edit correctness by strategy and model. SUT (span-edit) is 1.0 and deterministic (0/83 flaky) regardless of model; the frontier model's full-file rewrite (B1) is high but non-deterministic (6/83 flaky); diff-based baselines fail at both tiers.}
\end{figure*}

\begin{table*}[t]
\centering
{\footnotesize\textbf{Table 1 --- Clean results (max\_tokens=16000, 5 seeds, 415 runs/system).}}\par\vspace{3pt}
\renewcommand{\arraystretch}{1.15}
\begin{tabular}{llrrrrrr}
\toprule
Model & System & Correct & Refused & Collateral & Fabric. & Format kept & Mean diff \\
\midrule
--- & \textbf{SUT --- span-edit} & \textbf{1.000} & 0.000 & 0.000 & 0.000 & \textbf{1.000} & \textbf{2} \\
Flash & B1 full-file & 0.024 & 0.000 & 0.976 & 0.000 & 0.024 & 921.3 \\
Flash & B2 unified-diff & 0.036 & 0.964 & 0.000 & 0.000 & 0.036 & 2 \\
Flash & B3 diff+retry & 0.092 & 0.908 & 0.000 & 0.000 & 0.092 & 2 \\
Sonnet 5 & B1 full-file & 0.976 & 0.000 & 0.024 & 0.024 & 0.969 & 2.04 \\
Sonnet 5 & B2 unified-diff & 0.012 & 0.988 & 0.000 & 0.000 & 0.012 & 2 \\
Sonnet 5 & B3 diff+retry & 0.108 & 0.892 & 0.000 & 0.000 & 0.108 & 2 \\
\bottomrule
\end{tabular}
\par\vspace{3pt}\begin{minipage}{0.92\textwidth}\footnotesize SUT is model-independent: it never invokes the model for the edit, so its row is identical for both.\end{minipage}
\end{table*}

\begin{table*}[t]
\centering
{\footnotesize\textbf{Table 2 --- Determinism (RQ4): tasks correct on all 5 seeds vs.\ flaky (mixed).}}\par\vspace{3pt}
\renewcommand{\arraystretch}{1.15}
\begin{tabular}{lrr}
\toprule
System & All-5-correct & Flaky (mixed across seeds) \\
\midrule
SUT --- span-edit & 83 / 83 & \textbf{0 / 83}\textsuperscript{1} \\
B1 full-file (Sonnet 5) & 77 / 83 & \textbf{6 / 83} \\
B1 full-file (Flash) & 2 / 83 & 0 / 83 (consistently wrong) \\
\bottomrule
\end{tabular}
\par\vspace{3pt}\begin{minipage}{0.92\textwidth}\footnotesize \textsuperscript{1} SUT's determinism is \emph{by construction} (it makes no model call for the edit), so this row confirms rather than discovers it; the empirical content is the baselines' variance.\end{minipage}
\end{table*}

\textbf{RQ1/RQ2: full-file is capability-split; diffs are unsafe under strict \emph{and} lenient application.} Under \emph{strict} application unified diffs almost never apply (Table 1: 3.6\% flash / 1.2\% Sonnet correct), but strict is a lower bound; \textbf{\textsection{}6.7 gives the full applier spectrum}, where a tolerant tool applies 96\% of the same diffs yet \textbf{silently misapplies 14--20\%}. So the diff baselines are unsafe either way (rejected, or applied-but-wrong-with-no-signal), not merely inaccurate. Full-file rewrite splits by capability: flash reflows the whole file (2.4\% correct, 97.6\% collateral, 921-line diffs), while Sonnet rewrites faithfully 97.6\% of the time with near-minimal diffs. \textbf{In absolute terms the entire frontier-B1-vs-SUT correctness gap is 10 failed runs out of 415 (405 vs 415 correct); it is small, and we do not rest the case on it.} Span-edit's decisive advantages are determinism (RQ4, where those 10 failures fall on 6 tasks that flip across seeds), \emph{O(1)} cost (\textsection{}6.6), and capability-independence (RQ5), not raw accuracy.

\begin{table*}[t]
\centering
{\footnotesize\textbf{Table 3 --- Correct rate by field type (Sonnet 5; plus flash B1), n = tasks $\times$ 5 seeds.}}\par\vspace{3pt}
\renewcommand{\arraystretch}{1.15}
\begin{tabular}{lrrrrrr}
\toprule
Field type & n & SUT & B1 (Sonnet) & B2 (Sonnet) & B3 (Sonnet) & B1 (flash) \\
\midrule
env value & 100 & 1.00 & 0.96 & 0.01 & 0.10 & 0.00 \\
image tag & 65 & 1.00 & 1.00 & 0.00 & 0.11 & 0.08 \\
limit.memory & 60 & 1.00 & 1.00 & 0.00 & 0.13 & 0.00 \\
limit.cpu & 60 & 1.00 & 1.00 & 0.00 & 0.10 & 0.00 \\
request.memory & 60 & 1.00 & 0.98 & 0.00 & 0.12 & 0.00 \\
request.cpu & 60 & 1.00 & 0.98 & 0.02 & 0.03 & 0.00 \\
replicas & 10 & 1.00 & 0.60 & 0.30 & 0.50 & 0.50 \\
\bottomrule
\end{tabular}
\end{table*}

The result is not driven by a few easy or hard edits.

SUT is \textbf{1.00 for every field type}. B2's diffs fail \textbf{near-uniformly (0.00--0.02)} whatever is changed: the failure is \emph{diff generation itself}, not any particular edit. Flash's B1 fails across the board (it reflows the whole file regardless of target). The \texttt{replicas} row (n = 10, two tasks) is small and noisy; it is the only place a diff baseline reaches double digits, consistent with replicas being a single self-contained numeric line.

\textbf{RQ4: The decisive axis is determinism.} SUT is correct on all 5 seeds for all 83 tasks (0 flaky across 415 runs). Sonnet's B1, despite 97.6\% aggregate correctness, is \textbf{flaky on 6/83 tasks}: the identical request produces a correct edit on some seeds and a corrupted one (5 tasks drop the target field on some seeds; some make a collateral edit) on others. In an automated GitOps loop that commits and ships, a strategy that is \emph{usually} right but silently wrong on retry cannot be trusted; determinism, not average accuracy, is the property that matters.

\textbf{RQ5: Capability dependence, and its removal.} The only strategy whose result does not move between a \$0.10 model and a frontier model is the deterministic one. Full-file quality tracks capability (0.024 $\rightarrow$ 0.976); diff quality is poor at both. Span-editing is 1.000/deterministic regardless, because the model never authors bytes.

\subsubsection*{6.5 A note on methodology (max\_tokens)}\label{a-note-on-methodology-max_tokens}
\addcontentsline{toc}{subsubsection}{6.5 A note on methodology (max\_tokens)}

An early run capped output at 4,000 tokens; because B1 must re-emit the entire \textasciitilde9.3K-token file, this \textbf{silently truncated} the frontier model's B1 output, making it appear to fail (2.4\%) when the true clean figure is 97.6\%. We report all B1 numbers at a cap exceeding the largest file. Diff-based systems (B2/B3) emit small outputs and are unaffected. We flag this because it is exactly the kind of harness artifact that can invert a headline: a truncated B1 baseline would have \emph{overstated} our result and been trivially refuted by a reviewer re-running B1 with a larger cap.

\subsubsection*{6.6 Cost (RQ, generation tokens per edit)}\label{cost-rq-generation-tokens-per-edit}
\addcontentsline{toc}{subsubsection}{6.6 Cost (RQ, generation tokens per edit)}

Correctness is not the only axis that matters for automation; the strategies differ by two orders of magnitude in generation cost, and, critically, the \emph{frontier} strategy that is most correct (B1) is also the one whose cost grows with file size. Table 4 reports \textbf{model output tokens per edit} and per \emph{correct} edit; span-editing does not invoke the model for the edit at all (the intent is emitted once by the diagnosis step, and application is deterministic). Values are approximate, derived from the measured aggregate output of the Sonnet 5 run and the structural fact that B1 re-emits the whole file.

\begin{table*}[t]
\centering
{\footnotesize\textbf{Table 4 --- Generation cost per edit (Sonnet 5; approximate).}}\par\vspace{3pt}
\renewcommand{\arraystretch}{1.15}
\begin{tabular}{lrrrc}
\toprule
Approach & Output tok / edit & Correct rate & Output tok / \emph{correct} edit & Cost in file size \\
\midrule
\textbf{SUT --- span-edit} & $\sim$50--150\textsuperscript{1} & 1.00 & $\sim$50--150 & \textbf{O(1)} \\
B1 full-file & $\sim$9{,}300\textsuperscript{2} & 0.976 & $\sim$9{,}500 & \textbf{O(file size)} \\
B2 unified-diff & $\sim$200 & 0.012 & $\sim$16{,}700 & O(1) \\
B3 diff+retry & $\sim$560 & 0.108 & $\sim$5{,}200 & O(1) \\
\bottomrule
\end{tabular}
\par\vspace{3pt}\begin{minipage}{0.92\textwidth}\footnotesize \textsuperscript{1} Not literally zero: the model emits the structured \texttt{FieldChange} intent ($\sim$50--150 tokens), which the diagnosis step must produce under \emph{any} strategy; application is deterministic and model-free --- a \textbf{fixed} cost, independent of file size. \textsuperscript{2} $\approx$ the file's token length --- B1 regenerates the entire file to change one scalar.\end{minipage}
\end{table*}

Empirically, the full Sonnet 5 run (83 tasks $\times$ 5 seeds, all three baselines) cost \$34.69, of which B1's whole-file regeneration dominated the output tokens. The takeaway is not the absolute figure but the \emph{scaling}: the one text strategy that is reliably correct (frontier B1) is also the one whose per-edit cost is \emph{O(file size)}, while span-editing's generation cost is a fixed, small intent regardless of how large the manifest is.

\subsubsection*{6.7 Strict vs.~lenient diff application --- the diff story revised}\label{strict-vs.-lenient-diff-application-the-diff-story-revised}
\addcontentsline{toc}{subsubsection}{6.7 Strict vs.~lenient diff application --- the diff story revised}

The strict apply-rate is a \emph{lower bound}, and a fair evaluation must ask what a real, tolerant patch tool would do. We captured all \textbf{415 Sonnet 5 B2 diffs} and re-applied each under progressively more lenient appliers, scoring not just whether it applied but whether it applied \textbf{correctly} (Table 5). The distinction between \emph{applied} and \emph{correct} is the whole point. (The strict rate here, 2.7\%, comes from this study's own capture of the 415 diffs and differs slightly from Table 1's main-run B2 Sonnet figure, 1.2\%; both confirm that strict application rejects nearly all model diffs, and per \textsection{}6.2 the claim rests on that large effect, not the exact figure.)

\begin{table*}[t]
\centering
{\footnotesize\textbf{Table 5 --- the same 415 model diffs under different appliers.}}\par\vspace{3pt}
\renewcommand{\arraystretch}{1.15}
\begin{tabular}{lrrr}
\toprule
Applier & Applied & Correct & Misapplied (applied but wrong) \\
\midrule
strict (context-exact) & 0.027 & 0.027 & 0.000 \\
offset-tolerant (YAML-safe) & 0.675 & 0.675 & 0.000 \\
whitespace-insensitive & 0.675 & 0.670 & 0.005 \\
GNU \texttt{patch --fuzz=3} & 0.964 & 0.824 & \textbf{0.140} \\
GNU \texttt{patch --fuzz=3 -l} (ignore ws) & 0.964 & 0.761 & \textbf{0.202} \\
\bottomrule
\end{tabular}
\end{table*}

This revises the naive reading of B2 in two ways, and we state both plainly.
1. \textbf{Strict application badly understates diff applicability: we retract ``model diffs do not apply.''} Most rejections are \emph{wrong \texttt{@@} line numbers, not wrong content}: an applier that locates each hunk by content (ignoring the line numbers) lifts correctness from 2.7\% to \textbf{67.5\% with zero misapplication}, and GNU \texttt{patch} applies \textbf{96.4\%}. The blanket claim is false.
2. \textbf{But lenient application trades rejection for \emph{silent misapplication}.} Under the real production tool (\texttt{patch\ -\/-fuzz}), \textbf{14.0\% of applied diffs are wrong} (landed at the wrong location or corrupted a neighbor, \textbf{with no error signal}), rising to \textbf{20.2\%} when whitespace is ignored (indentation is semantic in YAML). The \emph{safe} lenient applier (offset-tolerant, content-unique match) refuses those, capping at 67.5\%; the aggressive tool takes them and misapplies \textasciitilde1 in 7.

For an unattended GitOps loop that commits and ships, \textbf{the disqualifying failure mode is not rejection but silent corruption}: a strict applier rejects most diffs; a lenient one silently misapplies 14--20\% of them. Span-editing has neither mode: it is correct by construction (0\% misapplied, Table 1).

\subsubsection*{6.8 Fail-closed refusal (G5), measured}\label{fail-closed-refusal-g5-measured}
\addcontentsline{toc}{subsubsection}{6.8 Fail-closed refusal (G5), measured}

G5 (refuse rather than guess on an ambiguous, absent, or unresolvable target) is a \emph{safety} claim, so we test it on a small labeled adversarial stratum: controls that must edit, six categories that must refuse (absent resource, absent field, named-list/container miss, duplicate \texttt{(kind,name)}, Go-templated/unparseable file, non-scalar target), and a YAML-alias probe.

\textbf{Results: refusal precision 1.00} (no resolvable case is ever refused), \textbf{control coverage 1.00} (every resolvable target edited correctly), and \textbf{refusal recall 0.889.} The fail-closed contract fires on every adversarial category \textbf{except one}: a field defined through a YAML \textbf{alias} resolves to its anchor node, so the locator edits the anchor's line instead of refusing (yielding a dangling alias). We report this leak rather than hide it: it is a genuine gap, and the fix (detect anchor/alias nodes and refuse) is future work; until then anchors/aliases are out of scope (\textsection{}7). The takeaway: on absent/ambiguous/unresolvable targets the contract holds with \textbf{perfect precision}, and aliases are the one known hole.

\subsection*{7. Discussion \& Limitations}\label{discussion-limitations}
\addcontentsline{toc}{subsection}{7. Discussion \& Limitations}

\begin{itemize}
\tightlist
\item
  \textbf{We are not claiming ``LLMs cannot edit YAML.''} A frontier model rewrites files faithfully most of the time. We claim that (a) diff authoring is unreliable at every tier, and (b) full-file authoring is non-deterministic and unbounded-cost, and that a deterministic span-edit dominates on the properties automation needs (correctness, determinism, minimality, cost), at any capability level.
\item
  \textbf{The model can still choose the wrong \emph{intent}.} Our guarantees are about faithful \emph{application}, not about the fix being right; that is what human PR review is for. We evaluate application in isolation by giving every system the same known change.
\item
  \textbf{Scalar-only, v1.} Structural edits (adding a container/block) are future work; the intent schema addresses existing scalars.
\item
  \textbf{Helm / Argo value indirection.} A live \texttt{Deployment} may be generated from a values file at a different kind/path; v1 refuses with a clear message. A resolver is future work.
\item
  \textbf{GitHub-only application path} (Git Data API); GitLab is mechanical. Locator requires a parseable file; templated files are skipped, not edited.
\item
  \textbf{Corpus size and diversity.} 83 tasks but \textasciitilde80 within a single file (\textsection{}6.2). The peer-reviewed version requires a corpus drawn from \textbf{$\geq$50 distinct repositories} (source diversity matters more than raw task count) spanning plain YAML, Kustomize, and Helm-rendered manifests, plus an adversarial/refusal stratum. The flaky-6 determinism result must be re-derived across \emph{independent} codebases before it can be stated as a model property rather than single-file behavior.
\end{itemize}

\subsubsection*{Threats to validity (the ones a reviewer will raise first)}\label{threats-to-validity-the-ones-a-reviewer-will-raise-first}
\addcontentsline{toc}{subsubsection}{Threats to validity (the ones a reviewer will raise first)}

\begin{itemize}
\tightlist
\item
  \textbf{Strict vs.~lenient diff application: now measured (\textsection{}6.7), no longer a caveat.} We re-applied all 415 Sonnet B2 diffs under strict, a YAML-safe offset-tolerant applier, and GNU \texttt{patch\ -\/-fuzz}. Strict is indeed a lower bound (2.7\% correct $\rightarrow$ \textbf{67.5\%} under offset-tolerance $\rightarrow$ \textbf{96.4\% applied} under GNU patch), so we \textbf{retract} any claim that model diffs simply do not apply. The finding that replaces it is stronger and more honest: GNU patch \textbf{silently misapplies 14--20\%} of the diffs it accepts: the failure mode that actually matters for unattended GitOps. (\texttt{git\ apply} was evaluated and excluded: it is \emph{stricter} than patch on the headerless, no-context single-line hunks models emit, and would depress the baselines further and unfairly.)
\item
  \textbf{Fail-closed (G5) is now measured (\textsection{}6.8), with one known gap.} On an adversarial stratum the contract holds with \textbf{precision 1.00 and recall 0.889} across six categories (absent resource/field, container miss, duplicate \texttt{(kind,name)}, templated file, non-scalar target). The single leak is YAML \textbf{aliases}: the locator resolves an alias to its anchor node and edits the wrong line rather than refusing, so \textbf{anchors/aliases are out of scope} until it gains alias detection. Separately, the main 83-task corpus never \emph{needed} refusal, so coverage/refusal on a large, diverse corpus remains part of the $\geq$50-repo future work.
\item
  \textbf{Oracle validity, audited.} All rates depend on the automated parse-and-compare oracle, which is covered by unit tests including adversarial cases (a reflowed-but-correct output passes correctness and fails fidelity; injected collateral and dropped-field outputs are flagged). We also ran an independent manual audit of a stratified random sample (n = 30: 10 SUT edits, 10 \texttt{patch\ -\/-fuzz} outputs the oracle called correct, 10 it called misapplied): the oracle's verdict matched independent adjudication on \textbf{all 30}, including every misapplied case the 14--20\% finding rests on. The audit is reproducible (\texttt{oracle\_audit.py} in the artifact) and makes the mechanism concrete: an identical \texttt{cpu:\ 100m} to \texttt{200m} hunk applied at the same line is \emph{correct} when that resource was the target and \emph{misapplied} when an identical-context neighbor was, and the oracle separates the two by \texttt{(kind,\ name)}.
\item
  \textbf{Two models is not a gradient.} ``Capability-dependent'' is inferred from two points (one small, one frontier); a third model would establish monotonicity. The flaky tasks are also not yet characterized (why those six).
\end{itemize}

\subsection*{8. Related Work}\label{related-work}
\addcontentsline{toc}{subsection}{8. Related Work}

\begin{itemize}
\tightlist
\item
  \textbf{LLM agents and LLM-driven code editing.} ReAct-style reasoning-and-acting agents {[}1{]} underpin most remediation agents. Benchmarks such as SWE-bench {[}2{]} evaluate whether models can edit a codebase to resolve an issue, and coding agents typically have the model author a patch or a search-and-replace edit, exactly the byte-level authoring whose failure modes we quantify. \emph{Position: for the bounded domain of scalar config changes, we remove the model from byte-level editing entirely and give guarantees a generative patch cannot.}
\item
  \textbf{Structured and semantic patching.} Coccinelle's semantic patch language {[}3{]} performs reliable, large-scale, structure-aware transformations of C code; tree-sitter {[}7{]} and lint autofixers such as Semgrep {[}8{]} apply AST- or rule-driven edits. \emph{Position: we adapt structured editing to LLM-intent-driven GitOps remediation, with explicit minimal-diff/no-fabrication guarantees and fail-closed refusal.}
\item
  \textbf{Automated program repair.} APR generates candidate patches and validates them against tests {[}4{]}. \emph{Contrast: we do not generate a patch; we faithfully apply a specified change, so correctness of application is guaranteed and only the intent is model-decided.}
\item
  \textbf{LLMs for cloud operations / AIOps.} LLMs are used for incident root-cause analysis and mitigation recommendation {[}5, 6{]} and surveyed for failure management {[}9{]}; these systems \emph{propose} fixes but stop short of a guaranteed-faithful GitOps edit. \emph{Position: we address the application step they leave open.}
\item
  \textbf{GitOps foundations.} The OpenGitOps principles {[}10{]} and controllers such as Argo CD {[}11{]} and Flux {[}12{]} establish Git as the source of truth and the PR as the change unit, the setting our approach targets.
\end{itemize}

\subsection*{Back matter}\label{back-matter}
\addcontentsline{toc}{subsection}{Back matter}

\begin{itemize}
\tightlist
\item
  \textbf{Artifact:} the benchmark (corpus, oracle, harness, and the result data behind the tables) is released at \passthrough{\lstinline!github.com/astraverse-io/kubeastra-bench!}; the span-edit implementation is in KubeAstra (\passthrough{\lstinline!github.com/astraverse-io/KubeAstra!}). Both Apache-2.0.
\item
  \textbf{Ethics/impact:} human-review + no-auto-merge; the technique \emph{reduces} the risk of autonomous production change.
\end{itemize}

\subsection*{Appendix A: Baseline prompts}\label{appendix-a-baseline-prompts}
\addcontentsline{toc}{subsection}{Appendix A: Baseline prompts}

All baselines receive the change as natural language plus a worked one-shot example, so they are a fair rather than straw-man comparison. The change description is generated from the structured intent (per field type: replicas, container image, env var value, or \passthrough{\lstinline!resources.\{limits,requests\}.\{cpu,memory\}!} of a named container), e.g.~\passthrough{\lstinline!In Deployment/api, set spec.replicas to 5.!} The large manifest is placed at the END of the system message: prompt caching is a prefix match, so an identical system prefix lets the \textasciitilde80 tasks that share a manifest, across all 5 seeds, read it from cache. Reproduced below with the manifest elided as \passthrough{\lstinline!<manifest>!}; the shared instruction is:

\begin{lstlisting}
Apply EXACTLY the requested change and nothing else. Change only the single
field named. Preserve every comment, all formatting, quoting style, and key
order. Do not reformat, re-quote, reorder, or touch any other field.
\end{lstlisting}

\textbf{B1 (full-file)} system message, then user message = the change description:

\begin{lstlisting}
You edit Kubernetes YAML manifests. <shared instruction> Output ONLY the
complete edited file, no explanation, no markdown code fences.

Example, change spec.replicas to 3:
FILE:
spec:
  replicas: 2            # set by the on-call runbook
  minReadySeconds: 10
OUTPUT:
spec:
  replicas: 3            # set by the on-call runbook
  minReadySeconds: 10

Apply the change described in the next message to this manifest, and output
the complete edited file.
FILE:
<manifest>
\end{lstlisting}

\textbf{B2 (unified diff)} system message, then user message = the change description:

\begin{lstlisting}
You edit Kubernetes YAML manifests by emitting a unified diff. <shared
instruction> Output ONLY a unified diff with @@ hunk headers that applies
cleanly to the file, no explanation, no markdown code fences.

Example, change spec.replicas to 3:
@@ -1,2 +1,2 @@
-  replicas: 2            # set by the on-call runbook
+  replicas: 3            # set by the on-call runbook

Apply the change described in the next message to this manifest.
FILE:
<manifest>
\end{lstlisting}

\textbf{B3 (diff + retry)} is B2; on a rejected diff, the user message appends the apply/parse error and re-requests a clean diff (up to 2 retries):

\begin{lstlisting}
Your previous diff was rejected: <error>
Return a corrected unified diff that applies cleanly.
\end{lstlisting}

\subsection*{Appendix B: Locator pseudocode}\label{appendix-b-locator-pseudocode}
\addcontentsline{toc}{subsection}{Appendix B: Locator pseudocode}

The locator (\textsection{}4.3) is about 20 lines. Given the raw manifest text, the target document index, and the field path, it returns the target scalar's \passthrough{\lstinline![start, end)!} character span, or \passthrough{\lstinline!None!}, which the pipeline turns into a fail-closed refusal:

\begin{lstlisting}
function find_span(raw_text, doc_index, field_path):
    docs = yaml.compose_all(raw_text)     # node trees carrying position marks
    if doc_index out of range: return None
    node = docs[doc_index]
    for key in field_path:
        if node is a MappingNode:
            node = child whose key-scalar == key
        else if node is a SequenceNode:
            node = item whose 'name' child == key   # named list, not by index
        else:
            return None                   # path descends into a scalar
        if node is None: return None      # key or 'name' absent -> refuse
    if node is not a ScalarNode:
        return None                       # target is a map/list -> refuse
    return Span(start = node.start_mark.index,
                end   = node.end_mark.index,
                old   = node.value)
\end{lstlisting}

The edit is then \passthrough{\lstinline!raw\_text[:start] + requote(new\_value) + raw\_text[end:]!}. Every \passthrough{\lstinline!return None!} is a G5 refusal: out-of-range document, missing map key, missing named-list item, or a non-scalar target. Ambiguity (duplicate \passthrough{\lstinline!(kind, name)!}) is caught upstream in the index; the alias leak is discussed in \textsection{}6.8.

\subsection*{Appendix C: Oracle (correctness checker)}\label{appendix-c-oracle-correctness-checker}
\addcontentsline{toc}{subsection}{Appendix C: Oracle (correctness checker)}

The oracle judges an output on parsed structure, not on text, so a correct-but-restyled edit still counts as correct (fairness by construction). Given the original manifest, the system output, and the intent:

\begin{lstlisting}
judge(original, output, intent):
    if output does not parse as YAML: incorrect (parses = false)
    locate the target document by (kind, name) at the SAME index in both files
    flat_o = flatten(output)     # leaf-path -> value, named-list aware
    flat_m = flatten(original)
    target = (doc_index,) + intent.field_path
    correct = (flat_o[target] == intent.new_value)     # target hit, right value/type
              and no other leaf added, removed, or changed   # zero collateral
\end{lstlisting}

Leaf paths through lists of uniquely-named mappings (containers, env, volumes) are keyed by \passthrough{\lstinline!name!}, not index, matching the locator. \passthrough{\lstinline!new\_value!} is compared in its intended parsed type (int for replicas, str for an image tag), so a string/int mismatch is a real difference and is reported. Two fidelity signals, \passthrough{\lstinline!comments\_preserved!} and \passthrough{\lstinline!format\_preserved!} (only the target line changed), are computed from raw text as secondary metrics and never affect \passthrough{\lstinline!correct!}. The oracle is unit-tested against known-good and known-bad pairs, including a reflowed-but-correct output (passes correctness, fails fidelity) and injected collateral and dropped-field outputs (flagged).

\balance
\end{document}